\def\vereq#1#2{\lower3pt\vbox{\baselineskip1.5pt \lineskip1.5pt
\ialign{$\m@th#1\hfill##\hfil$\crcr#2\crcr\sim\crcr}}}

\newcommand{\rhoi}{\rho_i}

\newcommand{\game}{\gamma_E}
\newcommand{\glin}{\gamma_{\mathrm{lin}}}

\newcommand{\Qgb}{Q_{\mathrm{gB}}}

\newcommand{\lamq}{\lambda_q}

\newcommand{\vthi}{v_{\mathrm{thi}}}

\newcommand{\exb}{{E\times B}}

\newcommand{\gammash}{\gamma_\mathrm{SH}}

\newcommand\ignore[1]{}

\documentclass[aps,prl,reprint,twocolumn,amsmath,amssymb,floatfix,superscriptaddress,nofootinbib,preprintnumbers]{revtex4-2}
\usepackage{graphicx}
\usepackage{dcolumn}
\usepackage{bm}
\usepackage[small,compact]{titlesec}
\usepackage{xcolor}

\usepackage{hyperref}
\hypersetup{
    unicode=false,           
    pdftoolbar=true,         
    pdfmenubar=true,         
    pdffitwindow=false,      
    pdfstartview={FitH},     
    pdfpagemode=UseOutlines, 
    pdftitle={Broadening of the Divertor Heat Flux Profile in High Confinement Tokamak Fusion Plasmas with Edge Pedestals Limited by Turbulence in DIII-D},    
    pdfauthor={D. R. Ernst},     
    pdfsubject={},    
    pdfcreator={},    
    pdfproducer={},   
    pdfkeywords={Turbulence, gyrokinetic, tokamak, non-ELM, H-Mode}, 
    pdfnewwindow=false,     
    colorlinks=true,        
    linkcolor=black,         
    citecolor=black,         
    filecolor=black,         
    urlcolor=blue          
}

\ifx\pdfoutput\undefined
\else
\DeclareGraphicsExtensions{.pdf,.png}
\fi%

\graphicspath{%
{.}%
{../../figs/}%
}

\makeatletter
\newcommand*{\balancecolsandclearpage}{%
  \close@column@grid
  \cleardoublepage
  \twocolumngrid
}
\makeatother

\newcommand\orcid[1]{\href{https://orcid.org/#1}{\includegraphics[height=0.8em]{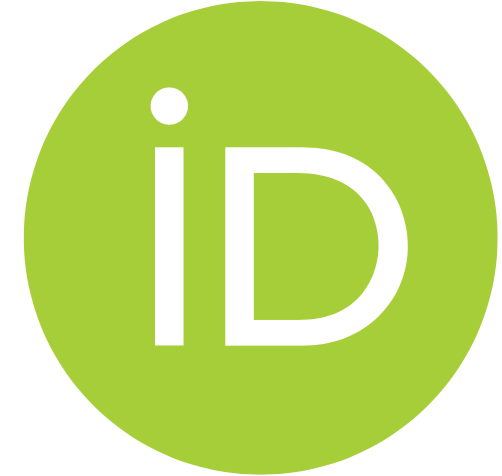}}}

\begin{document}
\titlespacing{\section}{0pt}{6pt}{6pt}

\preprint{
Phys. Rev. Lett. {\bf 132}, 235102 (2024).
\url{https://doi.org/10.1103/PhysRevLett.132.235102}
}

\title{Broadening of the Divertor Heat Flux Profile in High
  Confinement Tokamak Fusion Plasmas with Edge Pedestals Limited by
  Turbulence in DIII-D}

\author{D. R. Ernst\orcid{0000-0002-9577-2809}}
\affiliation{Massachusetts Institute of Technology, Cambridge, Massachusetts 02139, USA}
\email{dernst@psfc.mit.edu}
  
\author{A.~Bortolon\orcid{0000-0002-0094-0209}}
\affiliation{Princeton University Plasma Physics Laboratory, Princeton, New Jersey 08540, USA}
\author{C.~S.~Chang\orcid{0000-0002-3346-5731}}
\affiliation{Princeton University Plasma Physics Laboratory, Princeton, New Jersey 08540, USA}
\author{S.~Ku\orcid{0000-0002-9964-1208}}
\affiliation{Princeton University Plasma Physics Laboratory, Princeton, New Jersey 08540, USA}
\author{F.~Scotti\orcid{0000-0002-0196-9919}}
\affiliation{Lawrence Livermore National Laboratory, Livermore, California 94550, USA}
\author{H.~Q.~Wang\orcid{0000-0003-1920-2799}}
\affiliation{General Atomics, San Diego, California 92121, USA}
\author{Z.~Yan\orcid{0000-0002-5033-088X}}
\affiliation{University of Wisconsin, Madison, Wisconsin 53715, USA}
\author{Jie~Chen\orcid{0000-0002-4853-5341}}
\affiliation{University of California, Los Angeles, California  90095, USA}
\author{C.~Chrystal\orcid{0000-0003-3049-8658}}
\affiliation{General Atomics, San Diego, California 92121, USA} 
\author{F.~Glass\orcid{0000-0003-1101-1859}}
\affiliation{General Atomics, San Diego, California 92121, USA} 
\author{S.~Haskey\orcid{0000-0002-9978-6597}}
\affiliation{Princeton University Plasma Physics Laboratory, Princeton, New Jersey 08540, USA}
\author{R.~Hood\orcid{0000-0003-3789-3116}}
\affiliation{Sandia National Laboratory, Livermore, California 94551, USA}
\author{F.~Khabanov\orcid{0000-0002-6004-2005}}
\affiliation{University of Wisconsin, Madison, Wisconsin 53715, USA}
\author{F.~Laggner\orcid{0000-0003-1601-2973}}
\affiliation{North Carolina State University, Raleigh, North Carolina 27695, USA}
\author{C.~Lasnier\orcid{0000-0002-7109-2278}}
\affiliation{Lawrence Livermore National Laboratory, Livermore, California 94550, USA}
\author{G.~R.~McKee\orcid{0000-0002-2754-9816}}
\affiliation{University of Wisconsin, Madison, Wisconsin 53715, USA}
\author{T.~L.~Rhodes\orcid{0000-0002-8311-4892}}
\affiliation{University of California, Los Angeles, California  90095, USA}
\author{D.~Truong\orcid{0000-0002-8573-2539}}
\affiliation{Sandia National Laboratory, Livermore, California 94551, USA}
\author{J.~Watkins}
\affiliation{Sandia National Laboratory, Livermore, California 94551, USA}

\begin{abstract}
\noindent
Multimachine empirical scaling predicts an extremely narrow heat
exhaust layer in future high magnetic field tokamaks,  
producing high power densities that 
require mitigation.
In the experiments presented, the width of this exhaust layer
is nearly doubled  using actuators to increase turbulent transport in the
plasma edge. This is achieved in low collisionality, high confinement
edge pedestals with their gradients limited by turbulent transport instead of 
large-scale, coherent instabilities.
The exhaust heat flux profile width and divertor leg diffusive spreading both double as 
a high frequency band of turbulent fluctuations propagating in the electron diamagnetic direction
doubles in amplitude. The results are quantitatively reproduced in
electromagnetic XGC particle-in-cell simulations which
show the heat flux carried by electrons emerges to broaden the heat
flux profile, directly supported by Langmuir probe measurements.
\end{abstract}

\maketitle

\begin{figure}[h]
\includegraphics[width=\columnwidth]{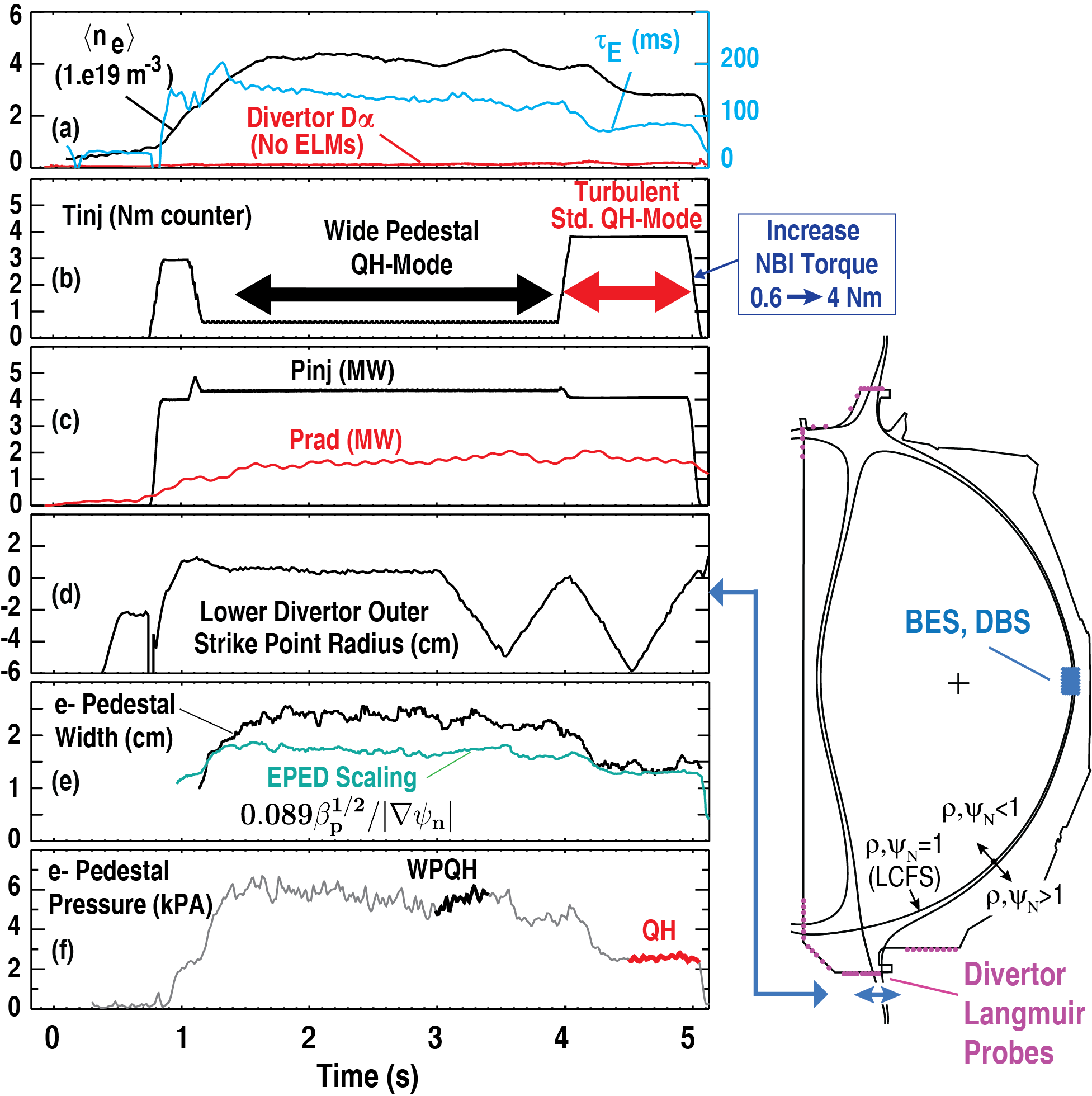}
\caption{
Evolution of (a) plasma density,  energy confinement 
time $\tau_E$, and divertor D$\alpha$ emission, 
(b) neutral beam torque $T_{\mathrm{inj}}$ counter to the plasma current, 
(c) injected neutral beam power $P_{\mathrm{inj}}$ and radiated power $P_{\mathrm{rad}}$, 
(d) lower divertor outer strike point major radius, 
(e) width of electron edge pedestal pressure relative to EPED scaling ($\psi_n$ is the 
poloidal magnetic flux enclosed normalized to its LCFS value), 
(f) electron pedestal top pressure (averaging time windows for analysis of 
WPQH/QH-mode phases shown).
\vspace{-1em}
 }
 \label{fig:wf}
\end{figure}

The tokamak is the main approach being pursued worldwide toward magnetic fusion energy
production.  It confines a plasma in a  toroidal magnetic field, typically several Tesla,
created by external solenoidal coils
with a helical twist added by the toroidally flowing plasma current. The
 magnetic field lines lie on nested, closed flux surfaces.
The fusion power produced is proportional to the square of the core 
plasma pressure.  Achieving high pressure requires good energy confinement, most often
achieved by creating a narrow edge transport barrier where 
densities and temperatures rise sharply just inside the last closed flux surface (LCFS).
Referred to as high confinement mode (H-mode), the edge transport barrier 
forms a ``pedestal'' which raises the entire pressure profile, approximately doubling 
the global energy confinement time.  The H-mode pedestal formation
is aided by using external coils to create a magnetic ``X point" at the LCFS
where the poloidal magnetic field vanishes (as shown later in Fig.~\ref{fig:wf}). 
The X point prevents magnetic field lines outside the LCFS from closing, causing 
them to be diverted to either side of the X point.  These open field lines then 
terminate on plates in the ``divertor'', which acts as a receptacle for the particle exhaust. 
Particles from the core which cross the LCFS stream down the open magnetic field lines 
to the divertor at the sound speed.
While the magnetic field lines are spread further apart in the divertor, the heat load 
from the particle exhaust is still concentrated in a narrow toroidal annulus on the divertor plates.
Further, in conventional H-modes, the pedestal pressure and plasma current density rise 
until an edge stability limit is reached, resulting in frequent, periodic edge localized 
mode (ELM) crashes which dump large bursts of energy and particles into the divertor.
Both the steady and transient components of this heat load
can reach tens of $\mathrm{MW/m^2}$, potentially  exceeding thermal 
stress limits and melting or eroding plasma facing components. 
Mitigating the plasma exhaust heat load remains one of the 
great challenges for the tokamak approach.
\begin{figure}[h]
\includegraphics[width=\columnwidth]{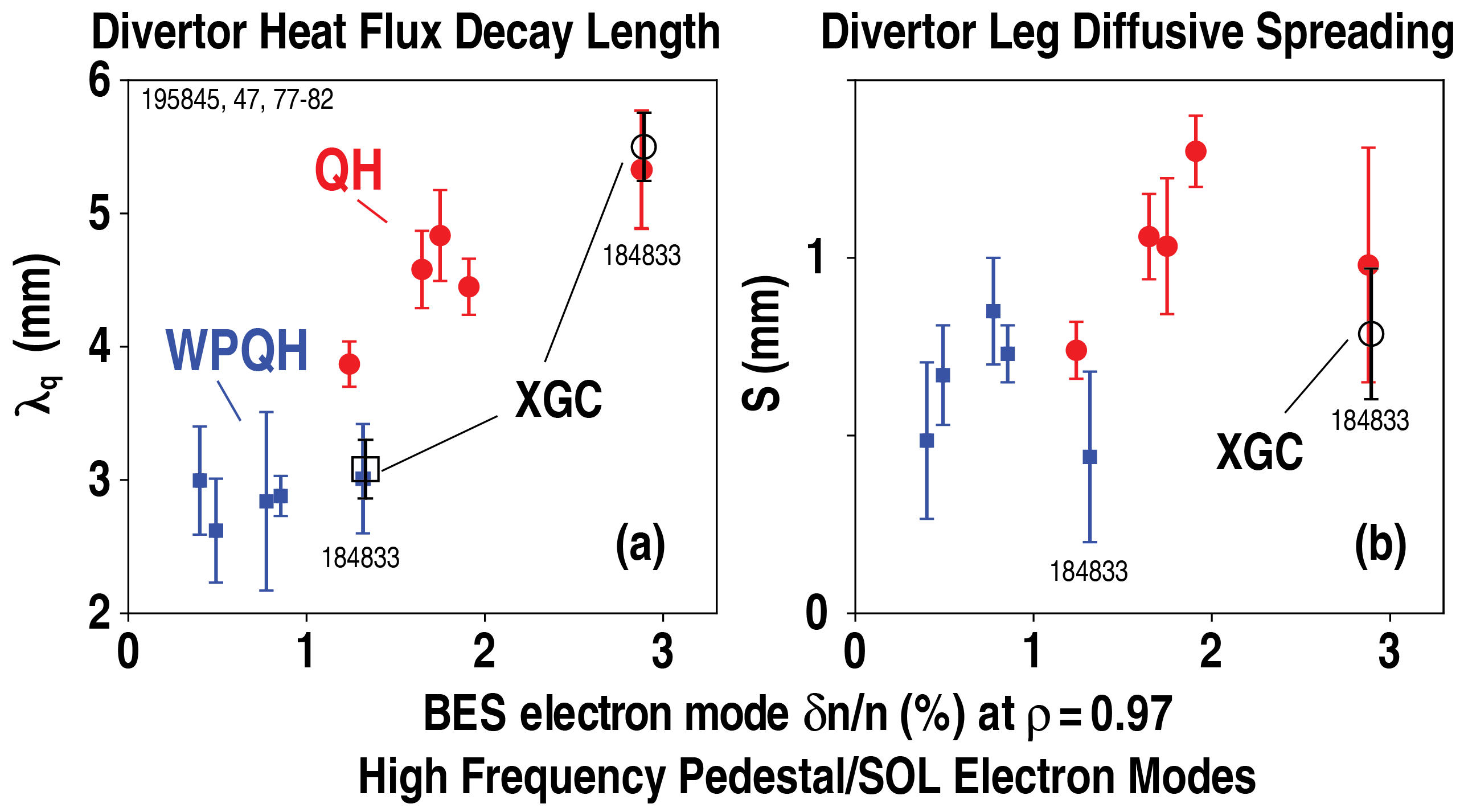}
\caption{(a) Measured $\lamq$ from Langmuir probes (LP)  
as a function of BES measured high-frequency electron mode 
density fluctuation amplitude $\delta n/n$ at $\rho=0.97$, and (b) 
Corresponding measured divertor leg diffusive spreading $S$.  Results from XGC simulations 
are shown in black open symbols using measured values from
No.~184833 for the $x$-coordinate (noise prevents determination of $S$ from XGC for WPQH). 
Error bars represent statistical standard deviations in the time averages.
Definitions of $\lamq$ and $S$ are given in the Appendix.
}
\label{fig:lamq_S_bes}
\end{figure}

\clearpage
\begin{figure}[h]
\includegraphics[width=\columnwidth]{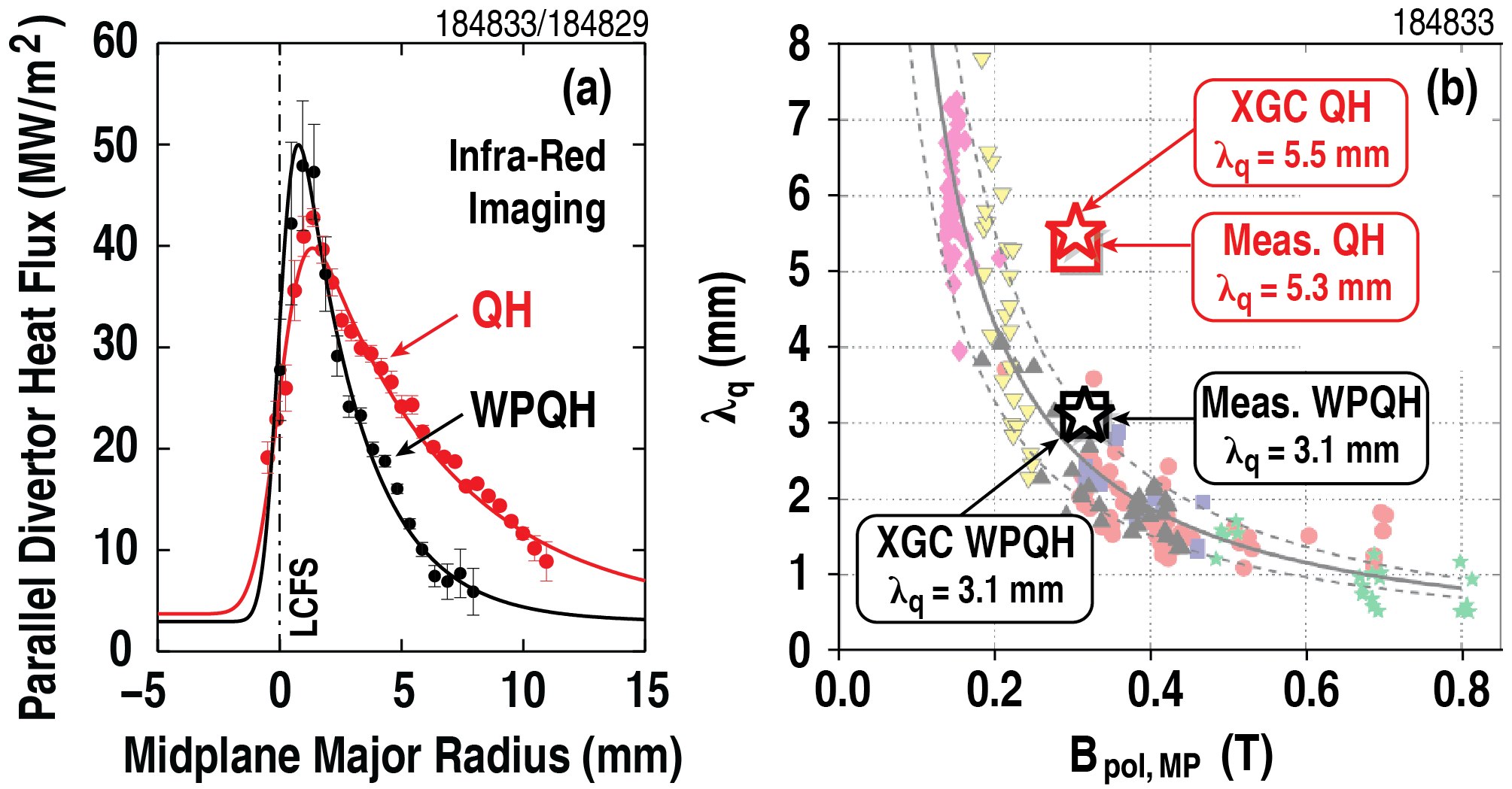}
\caption{(a) Divertor exhaust heat flux profiles measured by infrared thermography 
for turbulent QH-mode (No.~184833) and a matching WPQH-mode (No.~184829);
(b) Comparison of heat flux widths $\lambda_q$ from XGC
simulations with Langmuir probe measurements, against
the multimachine Eich scaling \cite{eich:2013} (DIII-D No.~184833).
}
\label{fig:xgc-eich}
\end{figure}

Macroscopic stability limits
the ratio of plasma pressure to magnetic pressure, so that stronger magnetic fields enable
access to higher pressures and thus higher fusion power. 
However, multimachine empirical scaling \cite{makowski:2012, eich:2013} would predict the exhaust heat flux width
depends inversely on (poloidal) magnetic field, so that stronger magnetic fields lead to more concentrated divertor heat loads.
For example, the scaling predicts a midplane width $\lamq\lesssim1$ mm for the heat exhaust layer
of the ITER tokamak, now under  construction.  The scaling is consistent with  
the heat flux being carried mainly by the ions, mediated by finite ion orbit width effects \cite{goldston:2012}. 

In this Letter we demonstrate experimentally that the divertor heat flux 
width (referred to the outer midplane) $\lamq$ increases with the intensity of high frequency 
turbulence propagating 
in the electron diamagnetic direction near the last closed flux surface.  
We show that for quiescent H-mode plasmas in the DIII-D tokamak where edge
turbulence is sufficiently strong,  $\lamq$ does not follow the empirical scaling, and can 
even increase favorably with poloidal magnetic field.
In these experiments we have used
actuators (varying applied toroidal torque and plasma current) to control the edge turbulence. 
Multiple diagnostic measurements of both edge 
turbulence and divertor heat flux profiles show that as the electron turbulence intensifies,
the fraction of the divertor heat flux carried by electrons increases and its profile broadens, broadening 
the total heat flux profile. 
The endpoints of the measured $\lamq$  range are 
quantitatively matched within measurement
uncertainty  by XGC gyrokinetic particle simulations, as summarized in
Figs.\ \ref{fig:lamq_S_bes} and \ref{fig:xgc-eich}.   
The turbulence produces large density fluctuations measured by beam emission spectroscopy (BES) \cite{McKee:1999}.  
Both $\lamq$ and $S$ increase with measured density fluctuation amplitude, so that the
total integral heat flux width  $\lambda_{\mathrm{int}}\approx \lambda_q + 1.64 \,S$  
also increases, as discussed in  the Appendix.
The mechanism for the broadening identified 
in these experiments lends plausibility to analogous XGC 
predictions of $\lamq  \sim 6$ mm for ITER \cite{chang:2017,chang:2021}.

In addition to the doubling of $\lamq$ shown, there are no ELMs  in these 
quiescent H-mode (QH-mode) plasmas,
which maintain steady high energy confinement with turbulent transport limiting 
pedestal gradients.
Broadening of $\lamq$ has also been observed in the quasi-continuous exhaust (QCE) or small ELM regime 
on the ASDEX Upgrade tokamak with increased fueling  \cite{faitsch:2021, faitsch:2023, eich:2020}, 
where it was associated with 
filamentary transport near the LCFS, 
thought to be driven by resistive  ballooning modes (RBM), 
though turbulence measurements providing direct evidence 
are not yet available.
The QCE regime operates with high edge collisionality  (ratio of trapped electron 
collision frequency to bounce frequency, $\nu_{*e}>10$)
which is beneficial for detaching the divertor (dissipating the heat flux by radiation), 
but it is not clear if ITER will be able to access
the necessary onset conditions for QCE
while maintaining the foreseen H-mode confinement. 
Future machines will operate at high temperatures and thus 
low collisionality at the pedestal top.
Our results are obtained with pedestal top collisionality  ($\nu_{*e}\sim 0.1$) similar to that expected in 
ITER regimes predicted to reach the goal of $Q=10$ fusion gain.
This low collisionality increases the pedestal gradient-driven bootstrap current, resulting in 
operation near the stability boundary for current-driven peeling modes rather than pressure 
gradient driven ballooning modes. 
Overall, we show the width of the exhaust layer in turbulence-limited QH-modes 
does not follow the empirical scaling with poloidal magnetic field $B_p$ (T) at the outer midplane, 
$\lamq^\mathrm{Eich}=0.63 B_p^{-1.19}$.
We argue that turbulence will be much stronger in future pedestals 
at higher magnetic field, which may naturally prevent ELMs while 
offering relief from the extreme steady divertor heat loads predicted by empirical scaling.

Figure \ref{fig:wf} shows the temporal evolution for the DIII-D discharge No.~184833
exhibiting the greatest $\lamq$ broadening, shown in Fig.\ \ref{fig:lamq_S_bes}.  
We first establish an H-mode
edge pedestal limited by turbulent transport without ELMs, exhibiting a high and wide pedestal pressure profile,
referred to as the wide pedestal quiescent H-mode  (WPQH-mode) regime \cite{burrell:2016, ernst:2018,
  burrell:2020, chen:2020}.  
Its pedestal width 
in normalized poloidal magnetic flux 
significantly exceeds
that predicted by EPED scaling \cite{snyder:2011, snyder:2012}, given by $0.089\beta_p^{1/2}$, 
where $\beta_p = 8\pi p/B_p^2$ is the pedestal top ratio of kinetic pressure to magnetic pressure, as shown.  
This regime is accessed through wall conditioning (boronization) to reduce collisionality, 
with low neutral beam injected (NBI) toroidal torque $<2$ Nm. 
During the period 4-5 s the counter-$I_p$ NBI torque is increased
from 0.6 to 4 Nm, provoking a back transition to 
a little explored variant of standard quiescent H-mode (QH-mode) which also exhibits only
broadband pedestal turbulence \cite{burrell:2005}, referred to here as
``turbulent QH-mode.''  More typical standard QH-mode pedestals are limited by
 low toroidal mode number $N=1-5$ Edge Harmonic 
Oscillations (EHOs). The energy confinement time,
pedestal pressure, and pedestal width step down significantly at this
transition, indicating increased turbulent transport.  
Pedestal radial profiles are shown in the Appendix.
The lower divertor heat flux profiles are measured by
fixed Langmuir probes and infrared (IR) camera imaging using strike point
sweeps in each phase.  Close agreement between Langmuir probes and IR thermography
is found when using the standard sheath heat flux transmission coefficient to relate
total parallel heat flux to Langmuir probe measured electron saturation current and electron temperature \cite{Stangeby:2000}. 
The comparison, together with detailed measurements for all discharges is given in the Appendix.

\begin{figure}[h]
\includegraphics[width=0.8\columnwidth]{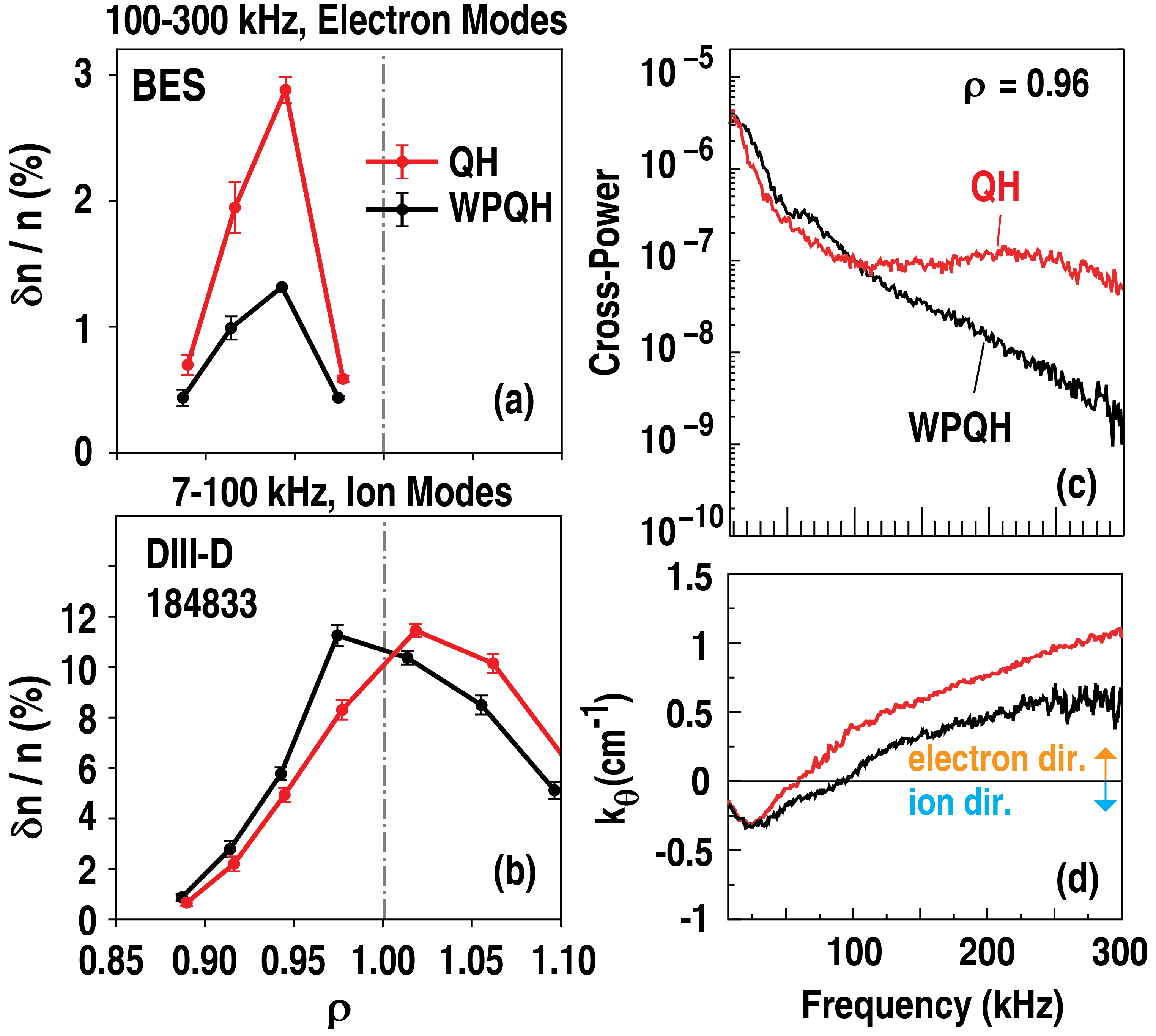} 
\caption{Pedestal radial profiles of normalized density fluctuation
  amplitude measured by BES, for
  WPQH-mode and Turbulent QH-mode phases in No.~184833, 
  for turbulence propagating in the lab
  frame (a) electron, and (b) ion diamagnetic directions, with (c) BES
  cross-power and (d) cross-phase ($k_{\theta}$ frequency spectra for
  vertically adjacent channels, showing modes in the ion (electron)
  diamagnetic direction at low (high) frequencies, at radial location
  $\rho=0.96$ corresponding to the peak $\delta n/n$ in (b).  Here $\rho$ is
  the minor radial coordinate given by the square root of toroidal magnetic 
  flux enclosed, normalized to its value at the LCFS.}
\label{fig:bes-profiles}
\end{figure}

\begin{figure}[h]
\includegraphics[width=\columnwidth]{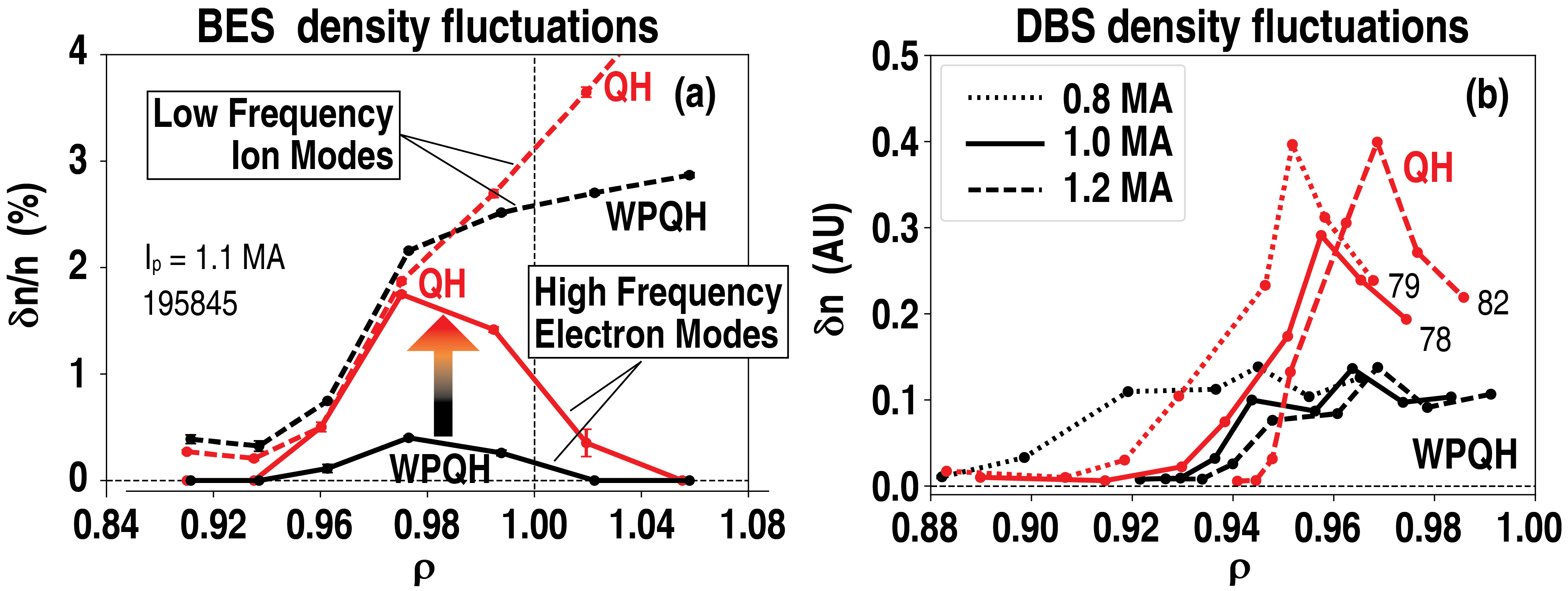}  
\caption{ (a) BES measurements of density fluctuations, showing lab
  frame ion-directed low frequency ($\lesssim 50-90$ kHz) modes at
  and high frequency ($\gtrsim 50-90$ kHz) electron-directed
  modes for both WPQH and QH phases, and (b) DBS
  measured $\delta n$ radial profiles for shorter wavelength modes in
  the range $k_{\theta} \sim 3-5$ cm$^{-1}$, $k_{\theta} \rho_s\sim
  1$, and toroidal mode number $N\sim 30-50$, generally corresponding
  to electron modes, comparing WPQH and QH phases for three plasma
  currents for discharges (discharges No.~195879, 78, 82) matched to 195845 in (a).}
\label{fig:bes-dbs-profiles}
\end{figure}

The increased pedestal energy transport and broadened $(\lamq, S)$ in Turbulent QH-mode are 
accompanied by a pronounced increase in high frequency density fluctuation amplitudes
measured by BES.
As shown in Figs.~\ref{fig:bes-profiles}(a), \ref{fig:bes-profiles}(b), two distinct
features are observed, a lower frequency band of fluctuations
propagating in the ion diamagnetic direction ($\sim$7 to 50--100 kHz),
and a higher frequency band of fluctuations above 50--100 kHz
propagating in the electron diamagnetic direction in the laboratory frame, 
as shown in Figs.~\ref{fig:bes-profiles}(c), ~\ref{fig:bes-profiles}(d). 
The low frequency ion-directed
fluctuations reach maximum intensity near the 
LCFS ($\rho\sim 1$) with
relatively little
change in intensity in the transition from WPQH-mode to QH-mode (the
calibration is not as reliable for $\rho>1$). On the other hand, the
measured high frequency electron-directed fluctuations in the pedestal
triple in amplitude in QH-mode relative to WPQH-mode.  
For reference, the increased
shear in the parallel flow due to the increased injected torque has
been shown to increase the drive for trapped electron modes (TEMs) in the QH-mode core
\cite{ernst:2016}.  
Measurements of the high frequency electron-directed fluctuations 
around the LCFS are not available for this XGC-simulated discharge, but are
available in the other more recent discharges (shown next). 
For a 1.1 MA case No.~195845 similar to No.~184833,
Fig.\ \ref{fig:bes-dbs-profiles}(a)  shows the low frequency fluctuations (in most
cases ion-directed)  change little in the transition from WPQH-mode to
QH-mode, even though there is a doubling of $\lambda_q$. This 
is an indication that the ion-directed fluctuation is not related 
to the widening of $\lambda_q$.
The low frequency amplitudes are comparable to
the higher frequency electron directed fluctuations and much weaker
than No.~184833.  The stronger electron fluctuations in QH-mode measured by BES are
confirmed in separate Doppler backscattering (DBS) measurements,
shown in Fig.~\ref{fig:bes-dbs-profiles}(b), selecting shorter poloidal
wavelengths than BES, $k_{\theta}
\sim 3-5$ cm$^{-1}$, $k_{\theta} \rho_s\sim 1$, and toroidal mode
numbers $N\sim 30-50$, where $\rho_s = (T_e/m_i)^{1/2}/\Omega_{ci}$ is the ion sound
gyroradius with $\Omega_{ci}=ZeB/m_ic$ the ion cyclotron frequency. The DBS measurements
show the same large relative change in density fluctuation
levels in the WPQH to QH-mode transition.
Unlike the discharge No.~184833, here the electron-direction fluctuations are measurable 
by BES at $\rho\sim1$ for several similar cases,  
extending outside the LCFS.

\begin{figure}[h]
\includegraphics[width=0.9\columnwidth]{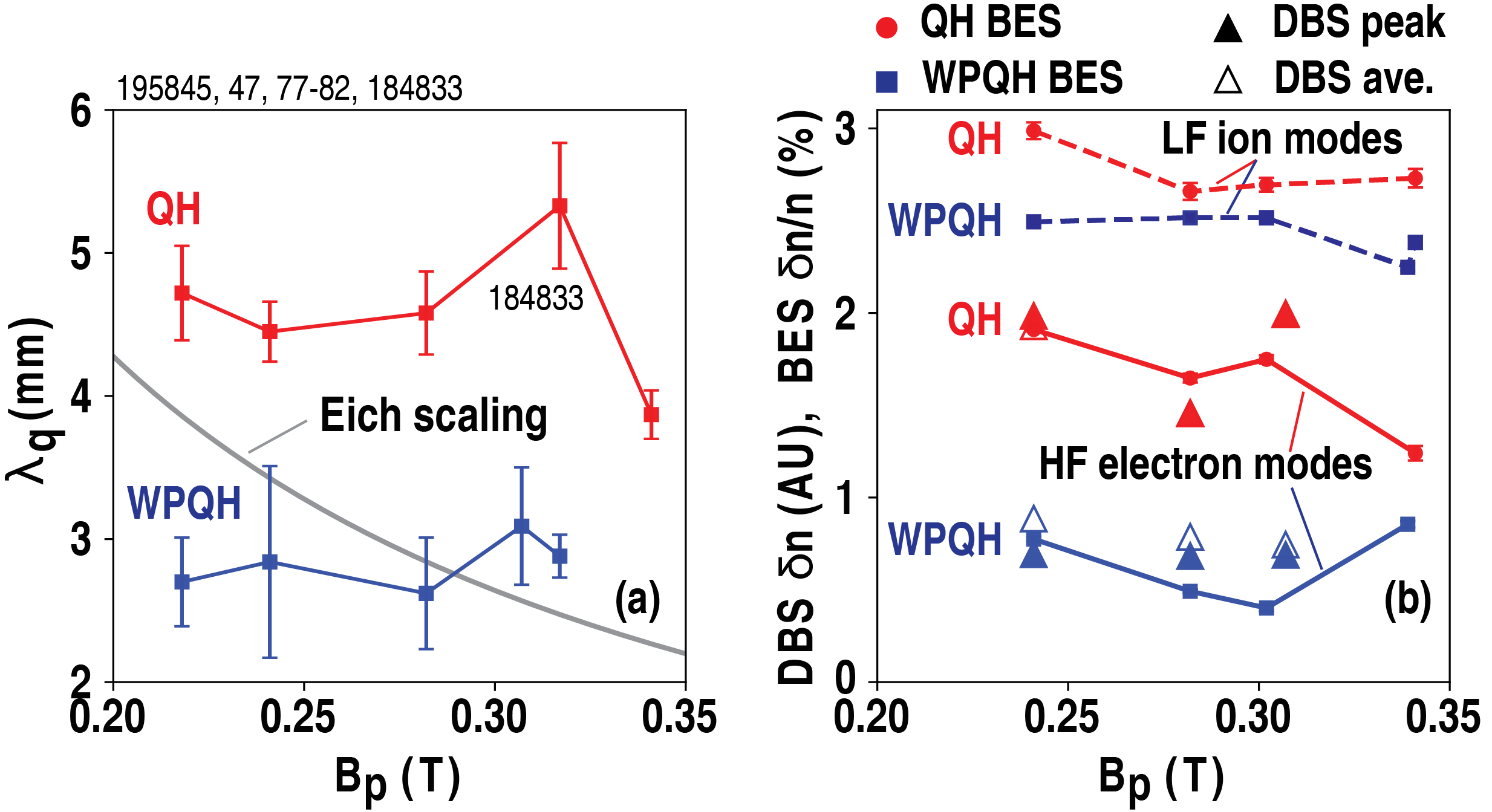}
\caption{Results of a recent plasma current scan: 
  (a) Measured $\lamq$ values from Langmuir probes as a function of outer midplane poloidal 
  magnetic field at the LCFS, $B_p$, 
  and (b) measured density fluctuation amplitudes from BES for high frequency
  ($\gtrsim 50-90$ kHz) modes in the electron direction at
  $\rho=0.97$, and for low frequency ($\lesssim 50-90$ kHz) modes
  propagating in the ion diamagnetic direction at $\rho=1.0$ (near the maximal amplitude), 
  together with DBS
  measurements of electron-directed $\delta n$ showing the peak value and the average of
  the outermost 5 radial points (see also
  Fig.\ \ref{fig:bes-dbs-profiles}).
  }
\label{fig:ipscan}
\end{figure}

These results are further confirmed in a plasma current scan (over the
range 0.7 to 1.3 MA).
While the $\lamq$ values measured in WPQH-mode are consistent with
 the multimachine Eich scaling \cite{eich:2013}, 
$\lamq$ in Turbulent QH-mode 
exhibits a more complex, favorable, nonmonotonic variation with $B_p$, as shown in Fig.~\ref{fig:ipscan}(a).
This nonmonotonic variation is reorganized as a monotonic increase with density fluctuation level in 
Fig.~\ref{fig:lamq_S_bes}.  The density fluctuation amplitudes, shown for high frequency electron modes
in Fig.~\ref{fig:ipscan}(b), show similar behavior to that of $\lamq$ in  Fig.~\ref{fig:ipscan}(a).
Density-weighted, radially line-integrated 
fluctuations of the fluctuating magnetic
field, $\delta B_R$, were measured by the radial
interferometer-polarimeter (RIP) \cite{jchen:2016}, and exhibit low and
high frequency bands similar to BES.  The high frequency band shows a
Doppler shift corresponding to the pedestal and is approximately four
times larger in magnitude than the low frequency band. Both bands
track the density fluctuation levels from BES, with the high frequency band doubling in amplitude as 
the BES high frequency band of density fluctuations doubles in amplitude.

\begin{figure}[h]
\includegraphics[width=\columnwidth]{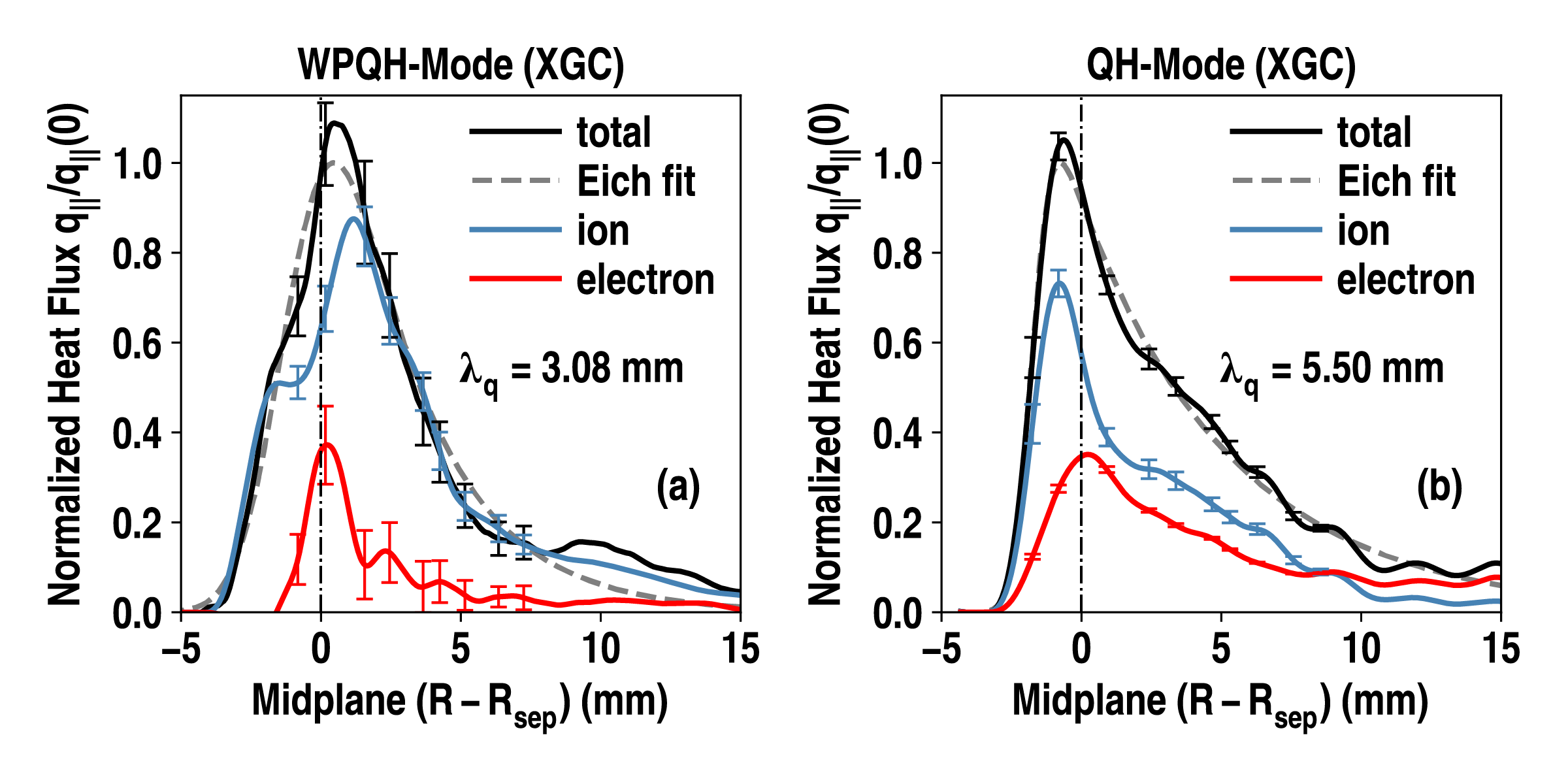}
\caption{XGC results for DIII-D No.~184833, showing the parallel heat
  flux at the last mesh point above the lower divertor plate, plotted
  against midplane major radius relative to the LCFS location,
  in (a) WPQH-mode and (b) QH-mode. Dashed curves show the fitted Eich
  function with corresponding $\lambda_q$.  The contributions of ions
  and electrons to the total heat flux are shown separately. Error bars represent
  the standard deviation on the time average.}
\label{fig:xgc-heatload}
\end{figure}

Both the WPQH-mode and QH-mode phases of discharge No.~184833 are simulated
with the XGC total-$f$ gyrokinetic particle code \cite{chang:2017},
using gyrokinetic ions and drift-kinetic electrons, and including Monte Carlo neutrals with
recycling coefficient 0.99, the carbon impurity density as measured, and electromagnetic effects.
Radial pedestal profiles used as input to XGC in are shown in the Appendix. 
Focusing on the electron mode turbulence, toroidal mode numbers $N=1-5$ are 
filtered out to avoid possible MHD related fast instabilities (this may artificially reduce the lowest 
frequency ion-directed turbulence amplitude). 
The overall results for $\lamq$ are shown in Fig.~\ref{fig:xgc-eich}, overlayed 
on the multimachine database \cite{eich:2013}.

We offer the following explanation for the measured  $\lamq$ 
broadening associated with the increased intensity of high frequency
electron turbulence.
The contributions in XGC simulations to the parallel heat flux at the
last mesh points above the divertor plates are shown in
Fig.\ \ref{fig:xgc-heatload} for the WPQH-mode (a) and QH-mode (b)
cases.  The WPQH-mode heat flux width $\lamq$ is consistent with the
Eich empirical scaling value, with most of the heat flux carried by
the ions. Its profile width is determined by ion magnetic drift velocity $v_{di}$. 
The Goldston heuristic model \cite{goldston:2012} yields the estimate 
$\lamq^\mathrm{Goldston}\sim v_{di} qR/v_{Ti}\sim \varepsilon\rho_\mathrm{pol,i}\propto B_\mathrm{pol}^{-1}$
where $qR$ is the field line length from midplane to divertor, $v_{Ti}$ 
is the ion thermal speed, $\varepsilon=r/R$ is the
inverse aspect ratio of the torus, and $\rho_\mathrm{pol,i}$ is the ion 
gyroradius using the poloidal magnetic field, 
which is consistent with the Eich scaling. The ion channel width was hypothesized 
to carry electron parallel heat flux \cite{goldston:2012}.
For the QH-mode, the XGC simulations reveal a greater and broader 
electron contribution to the heat
flux which emerges to broaden the total heat flux profile, 
associated with electron thermal transport from the higher temperature
pedestal region. The increased electron thermal transport can be 
expected to accompany the stronger electron turbulence
observed in measurements and simulations, which extends to the
LCFS.  TEM turbulence is known to most efficiently transport
electron thermal energy, particles, and impurities \cite{ernst:2016}.

\begin{figure}[ht]
\includegraphics[width=\columnwidth]{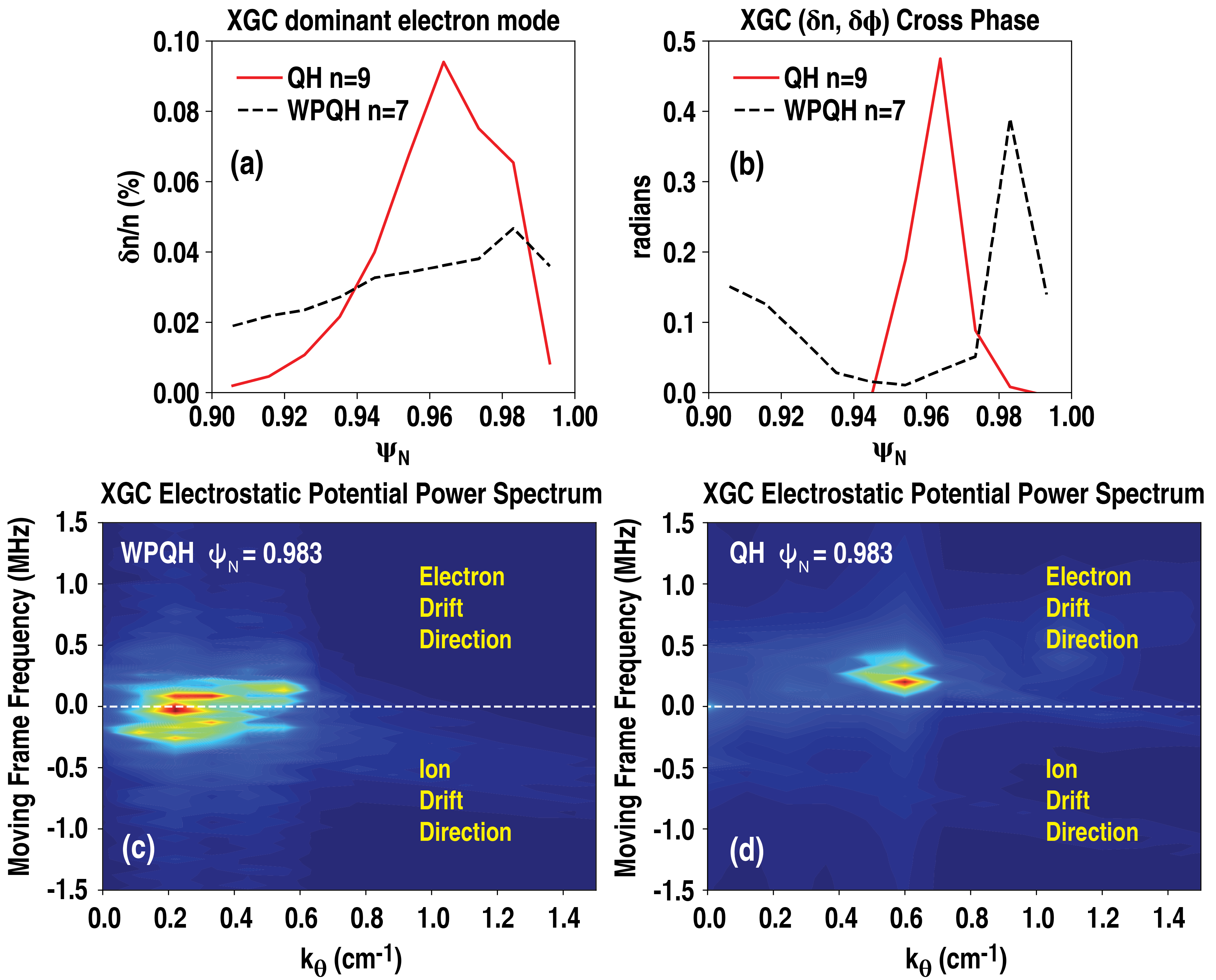} 
\caption{ (a) Density fluctuation level for the dominant electron mode
  in XGC simulations of No.~184833 for WPQH ($N=7$) and QH ($N=9$)
  phases, for modes propagating in the electron diamagnetic direction
  in the lab frame, and (b) the cross-phase between fluctuating
  density ($\delta n$) and electrostatic potential ($\delta \phi$) for
  the modes in (a); (c, d) XGC plasma frame frequency-wave number
  spectra corresponding to WPQH and QH phases, respectively, at
  $\psi_N=0.983$.  }
\label{fig:xgc-crossphase}
\end{figure}

The XGC simulations of toroidal mode number $N\! >5\!$ modes show overall consistency 
with the fluctuation measurements described above. 
Figure~\ref{fig:xgc-crossphase}(a) shows $\delta n/n$ for the dominant
toroidal mode number ($N$) fluctuations propagating in the electron
direction, for WPQH-mode ($N=7$) and QH-mode ($N=9$), indicating
that XGC simulations find a relative increase for this toroidal mode
similar to that measured in Figs.~\ref{fig:bes-profiles}(a) in the
steep gradient pedestal region.  
The corresponding cross-phase between density and
potential fluctuations $(\delta n, \delta \phi)$ from XGC is shown in
Fig.\ \ref{fig:xgc-crossphase}(b).  Noting that an ``adiabatic'' or
Boltzmann electron density response would be in phase with the electrostatic
potential, a strong nonadiabatic electron behavior (typical of the
trapped electron response) is indicated by the
non-zero cross-phase. This produces significant electron particle and thermal
transport, typical of TEM turbulence.  
From XGC, the lab frame frequency of the QH feature is 330 kHz,
while the poloidal wave number $k_{\theta}=0.63$ cm$^{-1}$, consistent
with the measurements in Figs.\ \ref{fig:bes-profiles}(c), \ref{fig:bes-profiles}(d).  In the
plasma frame moving with the $\exb$ velocity, the WPQH-mode case shows
both ion and electron modes spanning a range of wave numbers,
$k_{\theta}\sim 0.1 - 0.6$ cm$^{-1}$
(Fig.\ \ref{fig:xgc-crossphase}(c)), while in contrast, the QH-mode case shows
mainly electron modes with $k_{\theta}\sim 0.4 - 0.7$ cm$^{-1}$
(Fig.\ \ref{fig:xgc-crossphase}(d)). 

Initial collisionless estimates  using uniformly distributed Maxwellian 
tracer electrons in the region $0.98<\psi_N<1.0$ suggest that turbulent  
magnetic  fluctuations acting alone are less important than fluctuations 
in the $\exb$ velocity in causing these particles to hit the divertor. However, 
we have discovered a  strong synergy in which $\exb$ fluctuations nonlinearly 
amplify  electron transport by magnetic fluctuations. This new synergistic transport process 
will be explored in  future work. Homoclinic tangles \cite{evans:2004a}
are observed in the XGC simulations near the X point, and are more pronounced
in the QH-mode case, but subdominant.  

Finally, scaling arguments
suggest that in future machines, pedestal turbulent transport may be sufficiently
limiting to prevent ELMs \cite{ernst:2018, ernst:2022aps}, so that non-ELMing regimes naturally arise 
(see the Appendix for details). Intrinsically non-ELMing operating regimes, such as 
wide pedestal quiescent H-mode and quiescent
H-mode, offer a high-performance solution at relevant low pedestal top 
collisionalities, featuring
divertor heat flux profiles broadened by turbulence.

\begin{acknowledgments}

We thank Morgan Shafer (Oak Ridge National Laboratory) 
and Tom Osborne (General Atomics) for extensive comments and efforts 
during the  DIII-D internal review process; as well as 
both referees for their detailed and useful 
constructive comments. We would like to acknowledge
Keith Burrell (General Atomics), 
Florian Effenberg (Princeton Plasma Physics Laboratory),
Qiming Hu (Princeton Plasma Physics Laboratory),
Al Hyatt (General Atomics), 
Zeyu Li (General Atomics),
Adam McClean (Princeton Plasma Physics Laboratory), 
Tomas Odstr\v{c}il (General Atomics),
Theresa Wilks (Massachusetts Institute of Technology),
Guanying Yu (Univ. California, Davis),
and Lei Zeng (Univ. California Los Angeles) 
for contributions to the experiments featured.

This material is based upon work supported by the U.S. Department of
Energy, Office of Science, Office of Fusion Energy Sciences, using the
DIII-D National Fusion Facility, a DOE Office of Science user
facility, under Awards DE-FC02-04ER54698, DE-SC0014264,
DE-AC02-09CH11466, DE-SC0019004, DE-AC52-07NA27344, DE-NA0003525,
DE-FG02-08ER54999, and DE-SC0019352.  This research used resources of
the National Energy Research Scientific Computing Center (NERSC), a
U.S. Department of Energy Office of Science User Facility located at
Lawrence Berkeley National Laboratory, operated under Contract
No. DE-AC02-05CH11231. This research also used resources, via the 
INCITE program, of the Oak Ridge Leadership Computing Facility at 
the Oak Ridge National Laboratory, which is supported by the Office of Science 
of the U.S. Department of Energy under Contract No. DE-AC05-00OR22725.
Data included in figures and data required to reproduce simulation results are 
available from the author upon request.

This report was prepared as an account of work sponsored
by an agency of the United States Government. Neither the United
States Government nor any agency thereof, nor any of their employees,
makes any warranty, express or implied, or assumes any legal liability
or responsibility for the accuracy, completeness, or usefulness of any
information, apparatus, product, or process disclosed, or represents
that its use would not infringe privately owned rights. Reference
herein to any specific commercial product, process, or service by
trade name, trademark, manufacturer, or otherwise does not necessarily
constitute or imply its endorsement, recommendation, or favoring by
the United States Government or any agency thereof. The views and
opinions of authors expressed herein do not necessarily state or
reflect those of the United States Government or any agency thereof.

\end{acknowledgments}

\bibliographystyle{prsty}

\appendix*
\label{sec:appendix}
\textit{Eich Function.---} The integral heat flux profile width $\lambda_{\mathrm{int}}$ can be related to the Eich fit parameters $(\lambda_q, S)$ by $\lambda_{\mathrm{int}}=\int (q_{\|}(s) - q_{\mathrm{BG}}) ds/(q_{\|0})\approx \lambda_q + 1.64 \,S$   \cite{makowski:2012}. 
The Eich function  is given by 
\begin{equation}
q_{\|}(\bar{s})=\frac{q_{\|0}}{2} \exp \left[\left(\frac{S}{2 \lambda_q}\right)^2-\frac{\bar{s}}{\lambda_q f_x}\right] \operatorname{erfc}\left(\frac{S}{2 \lambda_q}-\frac{\bar{s}}{Sf_x}\right)+q_{\mathrm{BG}}
\end{equation}
where $\bar{s}=s-s_0=(R_{\mathrm{mp}}-R_{\mathrm{LCFS}})f_x$ is the radial coordinate at the divertor, referring to the departure in
midplane major radius from the LCFS, and 
$f_x = R_\mathrm{div}B_\mathrm{pol}^\mathrm{div}/R_\mathrm{mp}B_\mathrm{pol}^\mathrm{mp} \approx 5.3$
(for our cases) is the flux expansion effective area factor from midplane to divertor.  The two parameters $(\lambda_q, S)$ characterize the midplane exponential decay length and the Gaussian width, referred to the midplane, where $S$ represents competition between parallel and perpendicular heat transport between the X point and strike point \cite{eich:2013}.  The Eich function is the convolution of an exponential decay in the scrape-off layer and a Gaussian diffusive width characterized by S. Note the IR data for these experiments has insufficient coverage to constrain S, which is obtained from Langmuir probe measurements.

\begin{figure}[h]
\includegraphics[width=\columnwidth]{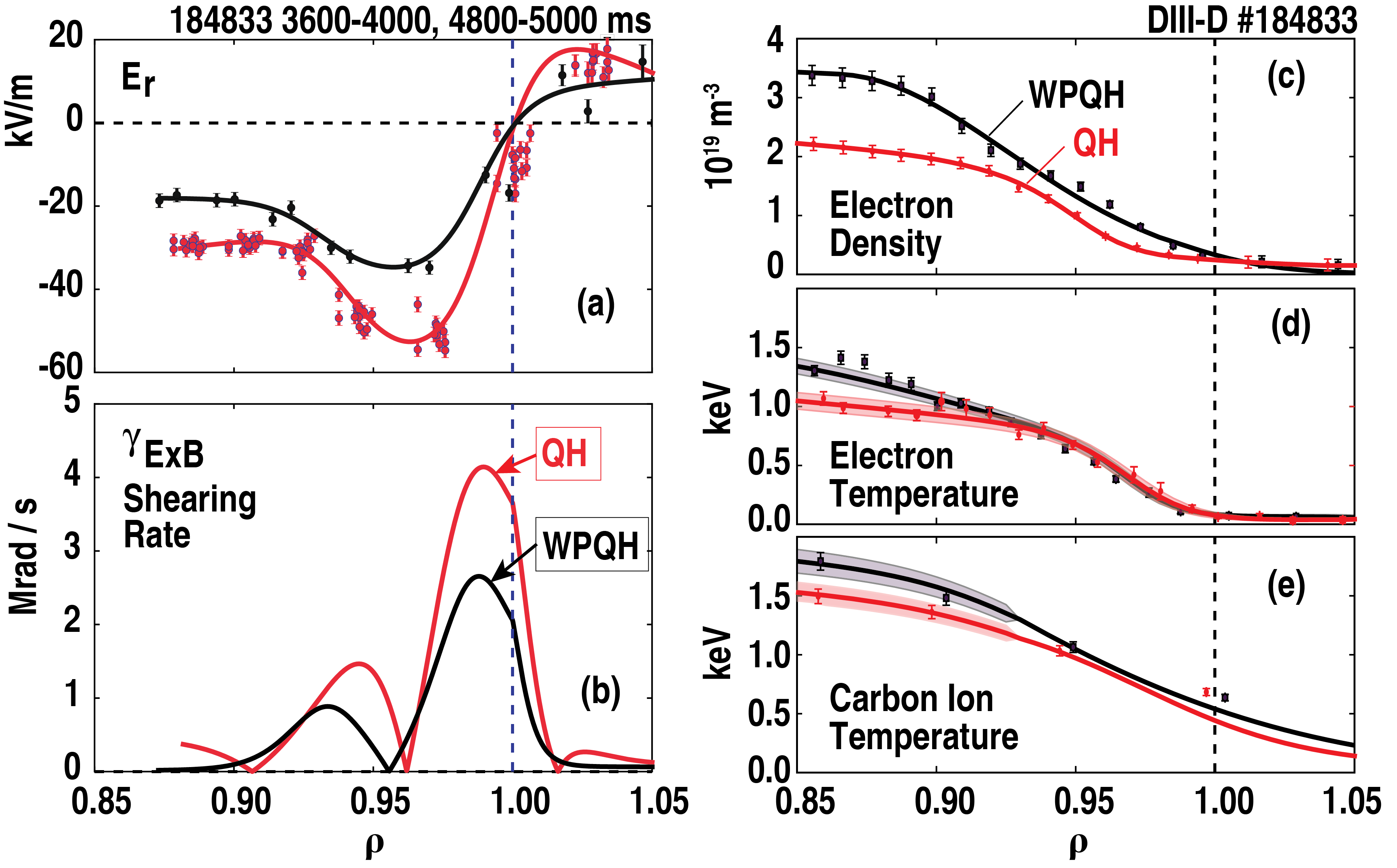}
\caption{Edge radial profiles for DIII-D No.~184833, comparing WPQH and QH-mode phases. (a) Radial electric field $E_r$, 
(b) Hahm-Burrell $\exb$ shearing rate, (c) electron density, (d) electron temperature, and (e) carbon ion temperature.
\vspace{-1em}
}
\label{fig:profiles}
\end{figure}

\textit{Profile Fits.---} Pedestal profiles, shown in Fig.~\ref{fig:profiles}, are constrained by
Langmuir probe measurements of separatrix electron temperature
$T_e^{\mathrm sep}$ and pedestal carbon and deuterium charge exchange
measurements, which show constant carbon concentration across the
pedestal. Equilibrium reconstructions of discharge No.~184833 in WPQH and QH phases, 
including the bootstrap current and realistic heating profiles from interpretative transport
simulations (TRANSP) are used as input to the XGC code.

\begin{figure}[h]
\includegraphics[width=0.6\columnwidth]{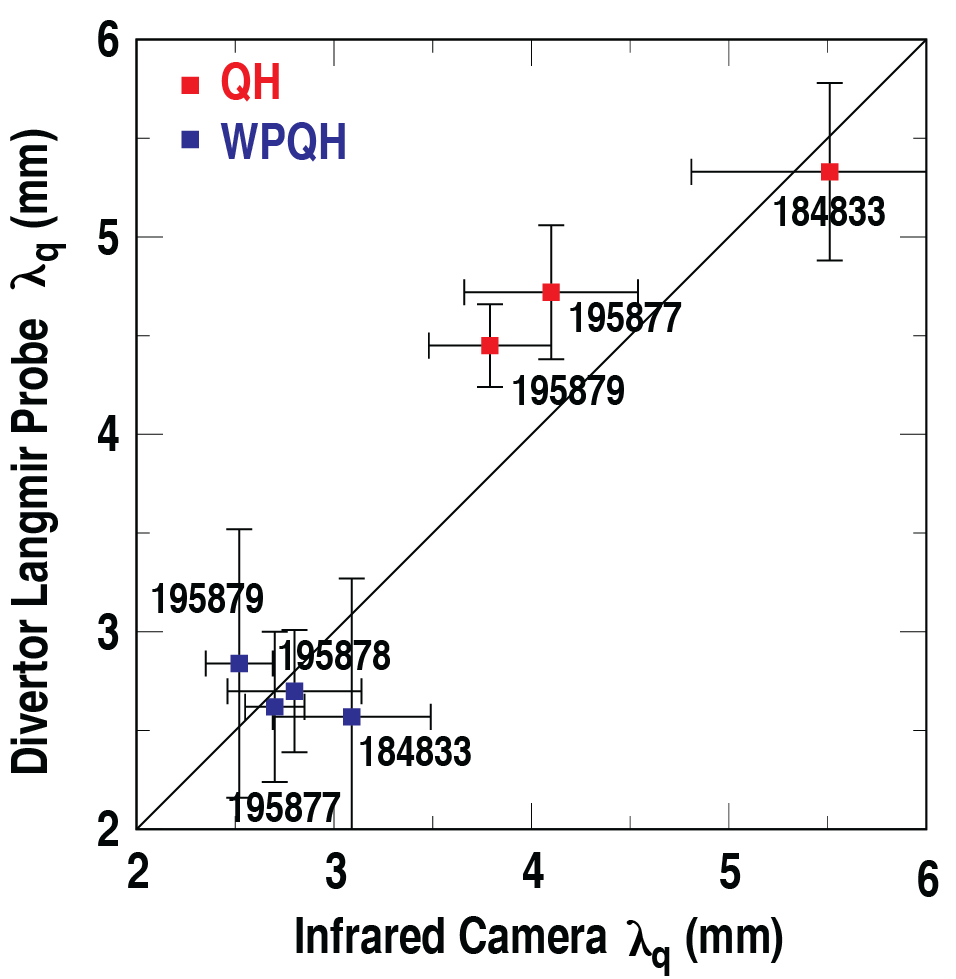}
\caption{Comparison of $\lamq$ inferred from Langmuir probe (using $\gammash=7$) and IR thermography.}
\label{fig:ir_lp_lamq}
\end{figure}

\begin{figure*}
\includegraphics[width=\textwidth]{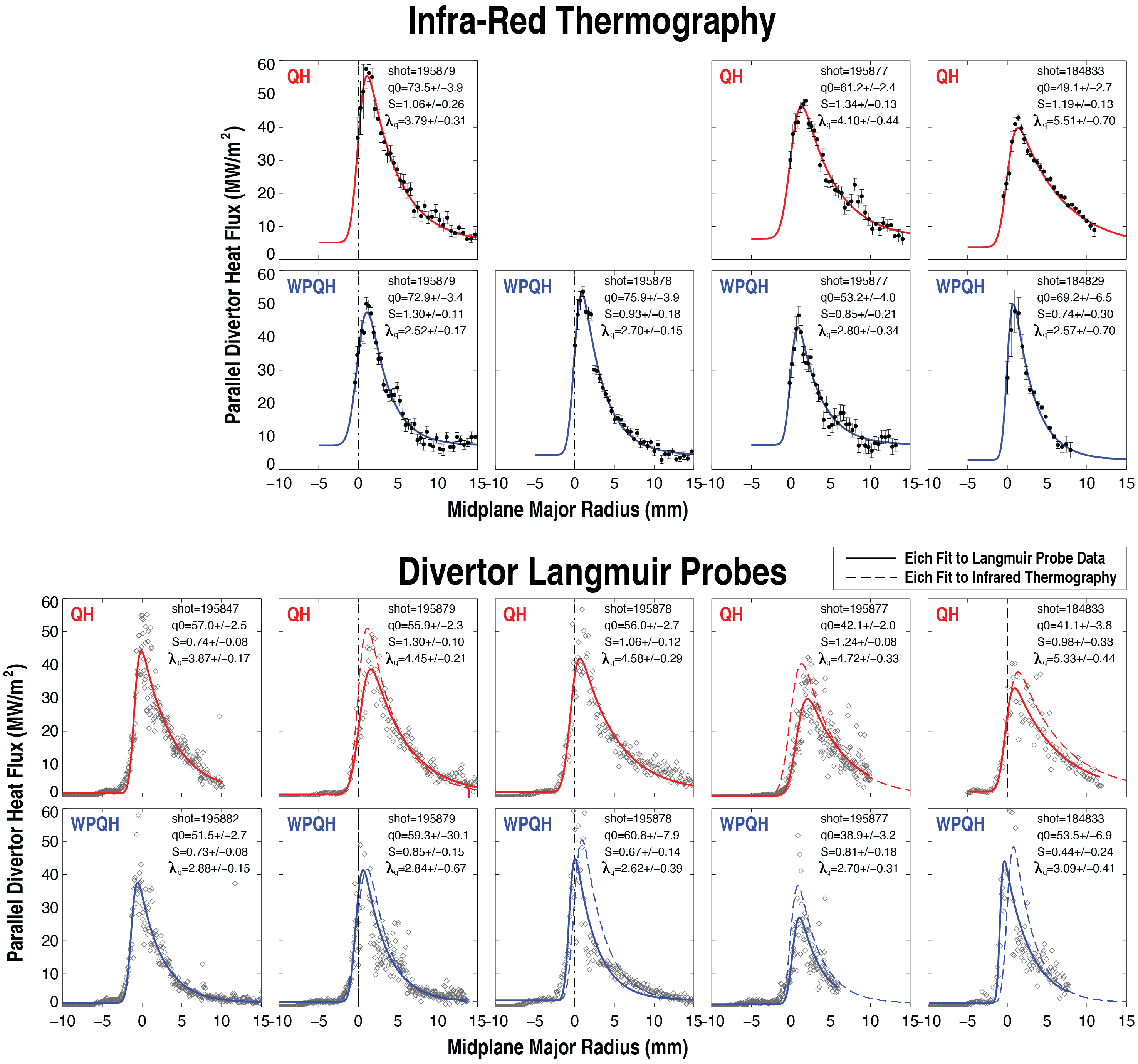}
\caption{Measured parallel heat flux profiles from IR thermography (above, for cases where measurement is available), and Langmuir probes (below) for the discharges in Fig.\ \ref{fig:xgc-eich}(b).  Solid lines show Eich fits with resulting paraminfraredeters given in the insets. The dashed lines overlay the Eich IR fit on Langmuir probe profiles after adjusting to match $q_\mathrm{BG}$. }
\label{fig:lp-ir-profiles}
\vspace{3em}
\end{figure*}

\textit{Divertor Heat Flux Measurements.---} Using the standard sheath heat transmission coefficient $\gammash=7$ \cite{Stangeby:2000} to relate total parallel heat flux from electrons and ions to Langmuir probe measured saturation current ($J_\mathrm{sat}$) and electron temperature ($T_e$) via  $q_{\|}^\mathrm{tot} = \gammash J_\mathrm{sat}T_e/e$, close agreement is obtained between Langmuir probe measured divertor heat flux profiles and IR thermography measured heat flux profiles, as shown in Fig.~\ref{fig:lp-ir-profiles}.  The IR measurements are shown for corresponding cases where available.  This agreement translates to  agreement within statistical measurement uncertainty in the $\lamq$ values inferred from Langmuir probes and IR, as shown in Fig.~\ref{fig:ir_lp_lamq}. Importantly, the $\lamq$ inferred from Langmuir probes is independent of the assumed $\gammash$.  Because Langmuir probes measure mainly electrons, this result can be viewed as evidence that the electron contribution to the divertor heat flux underlies the broadening of $\lamq$ measured by IR.

To clarify the issue of possible heat flux sharing between the upper and lower divertors, 
While the overall shape is double-null characterized by two
magnetic X points, for No.~184833 a -5 mm radial separation at the midplane between
their corresponding flux surfaces sends most of the heat flux to the
lower divertor \cite{petrie:2006}. In all of the other more recent discharges used in this study, the 
separation exceeds -10 mm, which well exceeds $\lamq$, so that heat flux sharing should not
affect the inference of $\lamq$.  

\textit{Scaling Pedestal Turbulent Transport to Future Machines.---}
Here we provide details on the scaling of pedestal turbulent transport with
ion gyroradius, which varies inversely with magnetic field. 
 Turbulent transport reduction by
sheared flows is predicted to scale with the normalized ion gyroradius,
$\rho_*=\rhoi/a$, where $a$ is the plasma minor radius
\cite{kotschenreuther:1996}. In future machines, $\rho_*$ is expected to 
be approximately 3 times smaller than in present tokamaks. 
The radial electric field $E_r$ in the
edge pedestal is $Zen_iE_r\sim dp_i/dr\sim p_i/\Delta$, where $Zen_i$
is the ion charge density, $p_i=n_iT_i$ is the ion pressure with $T_i$
the ion temperature, $r$ is the minor radius, and $\Delta$ is the
gradient scale length The $\exb$ shear rate for turbulence is then
$\game\sim B^{-1}dE_r/dr\sim E_r/(B\Delta)$ where $B$ is the magnetic
field.  Estimating the growth rate for ion scale drift-type
instabilities as $\glin\sim \vthi/\Delta$, where
$\vthi=(2T_i/m_i)^{1/2}$ is the ion thermal speed, the reduction of
turbulent transport by $\exb$ shear will depend on a parameter which
scales as $\game/\glin\sim \rho_i/\Delta \sim (a/\Delta)\rho_*$. The
ratio of pedestal turbulent heat flux $Q$ to the gyroBohm flux $\Qgb$
is predicted to increase asymptotically as a result of the decreased
$\exb$ shear at low $\rho_*$ according to
\cite{zhang:1992, kotschenreuther:2017, hatch:2018}
\abovedisplayskip=0.7em
\belowdisplayskip=0.7em
\[
{Q_{\hspace{2ex}}\over \Qgb}\sim
\left( {\game\over\glin}\right)^{-2}
\sim
\left({\Delta\over a}\right)^2 {1\over\rho_*^2}\sim {\beta_p^{2{\alpha_1}}\over \rho_*^2}
\label{eq:q_over_qgb}.
\]
Here $\beta_p = 8\pi p/B_p^2$ is the poloidal beta parameter, which
we have included by invoking the EPED scaling for the pedestal width
as limited by the onset of kinetic ballooning modes, $\Delta \sim
\beta_p^{\alpha_1}$, where $\alpha_1 \sim 0.5 - 0.75$
\cite{snyder:2011, snyder:2012}. Assuming an H-mode pedestal is
formed, pedestal turbulent transport could be approximately an order
of magnitude stronger in future high magnetic field machines as a result of the factor of
three reduction in $\rho_*$, assuming the pedestal width follows EPED scaling (KBM onset)  or 
a ``softer'' version of it with $\alpha_1<0.75$ due to other less virulent instabilities \cite{li:2024}.


\end{document}